\documentclass[sigconf]{acmart}

\usepackage{booktabs}
\usepackage{tabularx}
\usepackage{multirow}
\usepackage{microtype}
\usepackage{xspace}
\usepackage{enumitem}
\usepackage[ruled,vlined,linesnumbered]{algorithm2e}
\usepackage[most]{tcolorbox}
\usepackage[table]{xcolor}

\newcommand{\tablegroupcolor}{\rowcolor{black!6}}

\renewcommand\footnotetextcopyrightpermission[1]{}
\newcommand{\mypara}[1]{\vspace{0.3em}\noindent\textbf{#1.}\ }

\begin{document}

\title{Think Harder Where It Matters: Item-Level Adaptive Recursion for Sequential Recommendation}

\author{Zhanshuo Cao}
\authornote{Both authors contributed equally to this research.}
\affiliation{%
  \institution{Gaoling School of Artificial Intelligence, Renmin University of China}
  \city{Beijing}
  \country{China}
}
\email{caozhanshuo@ruc.edu.cn}

\author{Jiakai Tang}
\authornotemark[1]
\affiliation{%
  \institution{Gaoling School of Artificial Intelligence, Renmin University of China}
  \city{Beijing}
  \country{China}
}
\email{tangjiakai5704@ruc.edu.cn}

\author{Yuqi Zhou}
\affiliation{%
  \institution{Gaoling School of Artificial Intelligence, Renmin University of China}
  \city{Beijing}
  \country{China}
}
\email{yuqizhou@ruc.edu.cn}

\author{Sunhao Dai}
\affiliation{%
  \institution{Gaoling School of Artificial Intelligence, Renmin University of China}
  \city{Beijing}
  \country{China}
}
\email{sunhaodai@ruc.edu.cn}

\author{Jun Xu}
\authornote{Corresponding author.}
\affiliation{%
  \institution{Gaoling School of Artificial Intelligence, Renmin University of China}
  \city{Beijing}
  \country{China}
}
\email{junxu@ruc.edu.cn}

\renewcommand{\shortauthors}{Zhanshuo Cao et al.}

\begin{abstract}
Existing sequential recommenders usually allocate a fixed computational budget to every item in a user's interaction history, regardless of the item's contribution to the final prediction. We challenge this de facto design with two diagnostic findings. (1)~\textbf{Non-uniform computational need:} items within the same sequence vary drastically in how much computation they require. Most converge after shallow processing while a critical minority only stabilizes at full depth. (2)~\textbf{Suboptimality of fixed allocation:} given oracle access to the target item, a non-uniform depth assignment consistently outranks the fixed one, indicating that existing uniform computation both wastes capacity on already-converged items and sacrifices overall recommendation quality.

These observations motivate a shift from fixed-depth computation to \emph{item-level adaptive reasoning}. 
To this end, we propose \textbf{LoopRec}.
LoopRec recursively applies a single weight-shared block and introduces a lightweight gate at each recursion boundary that makes a binary exit-or-continue decision for every item independently. Items whose representations have sufficiently matured exit early, while those requiring deeper reasoning continue to subsequent recursions. This design converts the fixed global depth into a learned, per-item property without auxiliary depth labels, and is readily adaptable to diverse backbone architectures. Extensive experiments on public benchmarks with four backbone families demonstrate that LoopRec consistently outperforms matched fixed-depth counterparts. Further analysis reveals that LoopRec's learned allocation exhibits interpretable patterns aligned with our post-hoc oracle analysis: lower-frequency items and more recent interactions naturally receive deeper computation. Our code is available at \textcolor{blue}{\url{https://github.com/Elpmis117/LoopRec}}.

\end{abstract}


\maketitle

\section{Introduction}

Sequential recommendation models temporal patterns in user interaction histories to predict the next item of interest, supporting personalized ranking on e-commerce and online content platforms~\cite{kang2018sasrec,sun2019bert4rec,shin2024bsarec}. The dominant paradigm processes a user's interaction history through a fixed number of stacked layers, producing item representations that are then scored against candidates for next-item prediction. This paradigm has driven steady progress over the years, from attention-based models~\cite{kang2018sasrec,sun2019bert4rec,hou2022towards} to architectures incorporating frequency-domain inductive biases~\cite{shin2024bsarec,zhang2022attention,du2023frequency} and linear recurrences~\cite{yue2024linear,liu2024behavior,yang2026self}. Despite their architectural diversity, these methods share a common and largely unquestioned design choice: \emph{every item in the history receives exactly the same computational depth}.

We argue that this uniform allocation mismatches the nature of user behavior. A real-world interaction history is inherently heterogeneous~\cite{zhang2024explicit,rang2023heterogeneous}: it mixes routine repeat purchases that carry minimal predictive signal with rare, exploratory, or pivotal interactions that critically shape the user's evolving intent. Intuitively, these items should not demand the same amount of reasoning, yet current models are structurally incapable of differentiating between them, leading to suboptimal performance.

To move beyond intuition, we design two diagnostic experiments on a trained three-layer SASRec model (Section~\ref{sec:rethinking}). First, we find that $76.14\%$ of items already stabilize their contribution to the recommendation after a single layer, while $6.72\%$ remain sensitive through all three layers, and these two extremes frequently coexist within the same sequence. Second, a restricted oracle equipped with ground-truth target information finds the original fixed-depth allocation optimal in only $4.6\%$ of sequences, achieving up to $8.8\%$ relative Hit Rate improvement through flexible reassignment. These findings show that: \textbf{\emph{uniform depth both wastes capacity on items that have already converged and under-serves the critical minority that benefits from deeper computation}}. This motivates us to reframe sequential recommendation as a per-item computation allocation problem.

Realizing per-item adaptive computation in sequential recommendation, however, introduces two intertwined challenges. (1) \textbf{Representation alignment across depths:} since different items may exit at different depths, their output representations still need to reside in a shared feature space for downstream computation and scoring. Allowing depth to vary without careful architectural design can lead to misaligned representations that are incomparable at prediction time. (2)~\textbf{Progressive depth determination:} how much computation each item needs cannot rely on a single upfront decision, because an item's readiness only becomes apparent as it progressively absorbs context from surrounding items. Premature commitment risks either wasting computation on already-converged items or prematurely halting those that have not yet formed a reliable representation.

We propose \textbf{LoopRec}, a framework for sequential recommendation with item-level adaptive recursion. For \emph{representation alignment}, LoopRec uses a recursive architecture~\cite{dehghani2019universal,zhu2025scaling,tang2026loopctr,bae2025mor},
reusing a single weight-shared block across all depths. Since every recursion maps through the same parameters, items exiting at any depth naturally produce representations in the same feature space. For \emph{progressive depth determination}, LoopRec places a lightweight binary gate at each recursion boundary that evaluates each item's current state after absorbing the latest round of context, and decides whether it is ready to exit or should proceed to the next recursion. Items exit progressively as they converge, while those that benefit from deeper reasoning remain active. This design requires no auxiliary depth labels, introduces minimal additional parameters, and is readily adaptable to diverse backbone architectures.

We conduct extensive experiments on three public benchmarks with four different backbone families. LoopRec consistently outperforms the matched fixed-depth counterpart. Beyond accuracy gains, in-depth analysis of the learned routing reveals interpretable allocation patterns aligned with our post-hoc oracle analysis: the model naturally devotes deeper computation to lower-frequency items and more recent interactions, precisely the groups that an oracle would also allocate more computation to.

Our contributions are summarized as follows:
\begin{itemize}[leftmargin=1em]
  \item We use two complementary diagnostics to reveal variation in item-level computational demand and the limitations of fixed-depth processing in sequential recommendation.
  \item We propose LoopRec, which combines a shared recursive backbone with learned item-level stopping decisions to adapt computation depth within each interaction history.
  \item Experiments on three datasets and four backbones show consistent ranking improvements, with routing analyses revealing deeper computation for less frequent and more recent items.
\end{itemize}

\begin{figure}[t]
  \centering
  \captionsetup{skip=4pt}
  \includegraphics[width=0.98\linewidth]{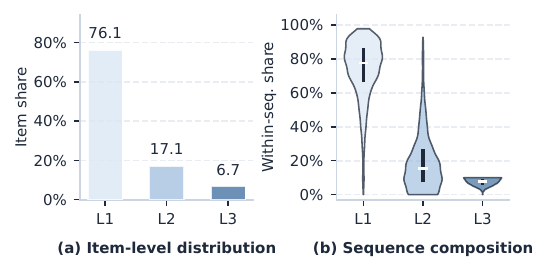}
  \caption{Item saturation depth $\ell_i^\star$ on Office. (a)~Item-level distribution across the test set. (b)~Per-sequence depth composition. Each violin shows the distribution of within-sequence shares, with dots and bars marking medians and IQRs.}
  \label{fig:rethinking_truncation_jaccard}
\end{figure}

\section{Diagnostic Analysis}
\label{sec:rethinking}

Before presenting LoopRec, we empirically probe the uniform-depth assumption with two diagnostic experiments on Amazon Office Products~\cite{hou2024bridging} using a three-layer SASRec~\cite{kang2018sasrec}. We investigate:
\begin{itemize}[leftmargin=1em, itemsep=2pt]
  \item \textbf{RQ1:} How is computational demand distributed across items within the same sequence under fixed-depth processing?
  \item \textbf{RQ2:} To what extent does the fixed-depth allocation affect recommendation quality relative to flexible per-item alternatives?
\end{itemize}

\subsection{Probe 1: Item-Level Saturation Depth}

To quantify the computational demand of individual items, we introduce the following diagnostic quantity:

\begin{tcolorbox}[enhanced, boxrule=0.4pt, colback=gray!3, colframe=gray!50, left=6pt, right=6pt, top=6pt, bottom=6pt, arc=1pt]
\small\textbf{Definition (Saturation Depth).}\quad For a trained $L$-layer model, the \emph{saturation depth} of item $i$ is the earliest layer $\ell$ at which its hidden representation can be frozen (i.e., excluded from all subsequent layer computations) without altering the model's final Top-$K$ recommendation set:
\begin{equation}
\ell_i^\star
=\min\!\left\{
\ell \in \{1,\dots,L\} \,:\,
\mathcal{R}_{\mathrm{trunc}}^{(i,\ell)}
=
\mathcal{R}_{\mathrm{full}}
\right\},
\end{equation}
where $\mathcal{R}_{\mathrm{full}}$ is the full-depth Top-$K$ set, and $\mathcal{R}_{\mathrm{trunc}}^{(i,\ell)}$ is the Top-$K$ set produced when item $i$'s representation is frozen after layer $\ell$ while all other positions continue through the remaining layers normally.
\end{tcolorbox}

\noindent Intuitively, a smaller $\ell_i^\star$ indicates that item $i$ contributes a ``settled'' signal early, while a larger $\ell_i^\star$ means it remains sensitive to deeper processing. We set $K{=}30$ and evaluate on all test sequences. We use this as a diagnostic statistic for the trained model. Additional details are in Appendix~\ref{app:fixed-depth-diagnostics}.

\paragraph{\textbf{Results.}}
Figure~\ref{fig:rethinking_truncation_jaccard}(a) reveals substantial variation in saturation depth: $76.14\%$ of items reach $\ell_i^\star=1$, and $93.28\%$ satisfy $\ell_i^\star \le 2$. Yet $6.72\%$ only saturate at full depth ($\ell_i^\star=3$). The sequence-level view in Figure~\ref{fig:rethinking_truncation_jaccard}(b) further shows that shallow- and deep-saturating items frequently coexist within the same history, with many sequences spanning all three saturation depths.

\begin{tcolorbox}[enhanced, boxrule=0.4pt, colback=blue!3, colframe=blue!40, left=6pt, right=6pt, top=4pt, bottom=4pt, arc=1pt]
\small\textbf{Finding 1: Non-uniform computational demand.}\quad Items within the same sequence exhibit highly heterogeneous computational demands. Most items converge after minimal processing, while a non-trivial minority requires full depth. Since both groups frequently coexist within the same sequence, no single fixed depth can simultaneously serve all items well.
\end{tcolorbox}

\subsection{Probe 2: Item-Level Depth Allocation}

Probe 1 establishes that items differ in how much computation they need, but does not tell us whether this mismatch actually hurts recommendation quality. To bridge this gap, we design a controlled experiment: keeping model parameters frozen, we ask whether a per-item flexible depth assignment can produce better rankings than the original fixed allocation.

Specifically, we construct a counterfactual search over the five most recent items in each test sequence. Each of these items is allowed to exit after depth 1, 2, or 3, while earlier positions retain the original full-depth pass. The original fixed-depth assignment is included as a candidate. An offline oracle with access to the ground-truth target then selects the assignment that achieves the best target rank. Construction details are in Appendix~\ref{app:fixed-depth-diagnostics}.

\begin{table}[t]
  \centering
  \caption{Rank improvements and relative tail computation for the sequences in Figure~\ref{fig:rethinking_tail5_oracle}(b), normalized by the fixed-depth allocation.}
  \label{tab:rethinking_tail5_cases}
  \small
  \setlength{\tabcolsep}{4pt}
  \renewcommand{\arraystretch}{1.08}
  \begin{tabular}{@{}lccccc@{}}
    \toprule
    Sequence & Seq.\,1 & Seq.\,2 & Seq.\,3 & Seq.\,4 & Seq.\,5 \\
    \midrule
    Rank & $24{\to}13$ & $7{\to}4$ & $7{\to}5$ & $12{\to}10$ & $19{\to}18$ \\
    Gain & $+11$ & $+3$ & $+2$ & $+2$ & $+1$ \\
    Rel.\ tail comp. & $40.0\%$ & $66.7\%$ & $53.3\%$ & $60.0\%$ & $60.0\%$ \\
    \bottomrule
  \end{tabular}
\end{table}

\begin{figure}[!t]
  \centering
  \captionsetup{skip=4pt}
  \includegraphics[width=\linewidth,trim=0 4bp 0 0,clip]{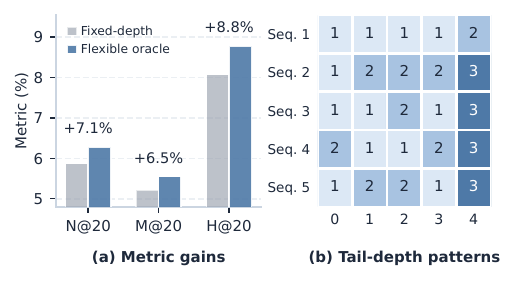}
  \caption{Restricted tail-5 oracle on Office. (a)~Relative gains of the flexible oracle over the fixed-depth baseline. (b)~Oracle-selected tail depths for sample sequences. Numbers denote depths, and columns 0--4 are the five most recent positions.}
  \label{fig:rethinking_tail5_oracle}
\end{figure}

\paragraph{\textbf{Results.}}
Figure~\ref{fig:rethinking_tail5_oracle} and Table~\ref{tab:rethinking_tail5_cases} present the oracle's findings. (a)~Aggregated over the test set, the flexible oracle improves over the fixed-depth baseline by $7.1\%$ on NDCG@20, $6.5\%$ on MRR@20, and $8.8\%$ on HR@20. (b)~Concrete examples show that the oracle-selected assignments achieve rank improvements while often consuming \emph{less} total tail-side computation than the fixed baseline. In addition, the fixed-depth assignment is the oracle's preferred choice in only $4.6\%$ of test sequences. In the remaining $95.4\%$, a non-uniform allocation ranks the target higher.

\begin{tcolorbox}[enhanced, boxrule=0.4pt, colback=blue!3, colframe=blue!40, left=6pt, right=6pt, top=4pt, bottom=4pt, arc=1pt]
\small\textbf{Finding 2: Suboptimality of fixed allocation.}\quad The fixed-depth allocation is locally suboptimal in the overwhelming majority ($95.4\%$) of sequences. A flexible per-item assignment improves ranking quality while often reducing total computation, demonstrating that uniform depth actively leaves performance on the table.
\end{tcolorbox}

\vspace{0.5em}
These two findings establish a strong empirical case for treating depth as a \emph{per-item computation allocation variable} in sequential recommendation, motivating the design of LoopRec.

\section{The Proposed LoopRec}
\label{sec:method}

To address the inefficiency of fixed-depth computation in sequential recommendation, we propose LoopRec, an adaptive routing framework built on a single shared backbone. LoopRec recursively reuses the same backbone block and lets a depth router assign different history positions to different recursion depths, so that computation is dynamically allocated item by item while the standard next-item prediction objective remains unchanged. Sec.~\ref{sec:task} formulates the task, Sec.~\ref{sec:LoopRec} presents depth-routed recursion, and Sec.~\ref{sec:objective} describes the training objective.

\subsection{Task Formulation}
\label{sec:task}

We consider the standard next-item prediction task in sequential
recommendation over an item set $\mathcal{V}$. Each training instance is an interaction sequence $\mathbf{s}=[v_1,\dots,v_n]$, where $v_t\in\mathcal{V}$ and the items are ordered by interaction time.
Given the prefix $\mathbf{s}_{1:t}=[v_1,\dots,v_t]$, the goal is to predict the next item $v_{t+1}$, where $1\le t<n$.

Each item $v_t$ is mapped to an input token
$\mathbf{x}_t=\phi(\mathbf{e}_{v_t}+\mathbf{p}_t)$, where
$\mathbf{e}_{v_t}\in\mathbb{R}^{d}$ and
$\mathbf{p}_t\in\mathbb{R}^{d}$ are the item and positional embeddings.
Stacking the positions yields
$\mathbf{X}\in\mathbb{R}^{T\times d}$ for the padded sequence length
$T$.
For each prefix $\mathbf{s}_{1:t}$, the model outputs a probability distribution over candidate items, denoted by
$p(v\mid \mathbf{s}_{1:t})$. Given the training set $\mathcal{D}$ of prefix--target pairs, the recommendation objective is the standard cross-entropy loss:
\begin{equation}
\mathcal{L}_{\mathrm{rec}}
=-\frac{1}{|\mathcal{D}|}\sum_{(\mathbf{s}_{1:t},\,v_{t+1})\in\mathcal{D}}
\log p(v_{t+1}\mid \mathbf{s}_{1:t}).
\end{equation}

\subsection{LoopRec Architecture}
\label{sec:LoopRec}

Section~\ref{sec:rethinking} shows that different historical items within the same sequence can require different amounts of computation, making uniform fixed-depth processing restrictive for sequential recommendation. Motivated by this observation, LoopRec allocates computation depth at the item level while keeping the standard
next-item prediction objective unchanged. For each position $t$, let $\mathbf{e}_{v_t}\in\mathbb{R}^{d}$ and $\mathbf{p}_t\in\mathbb{R}^{d}$ denote the item and positional embeddings, and let $\mathbf{x}_t=\phi(\mathbf{e}_{v_t}+\mathbf{p}_t)$ be the corresponding input token. Stacking these tokens gives the padded input sequence $\mathbf{X}=[\mathbf{x}_1,\dots,\mathbf{x}_T]\in\mathbb{R}^{T\times d}$.

\begin{figure}[!t]
  \centering
  \includegraphics[width=\linewidth]{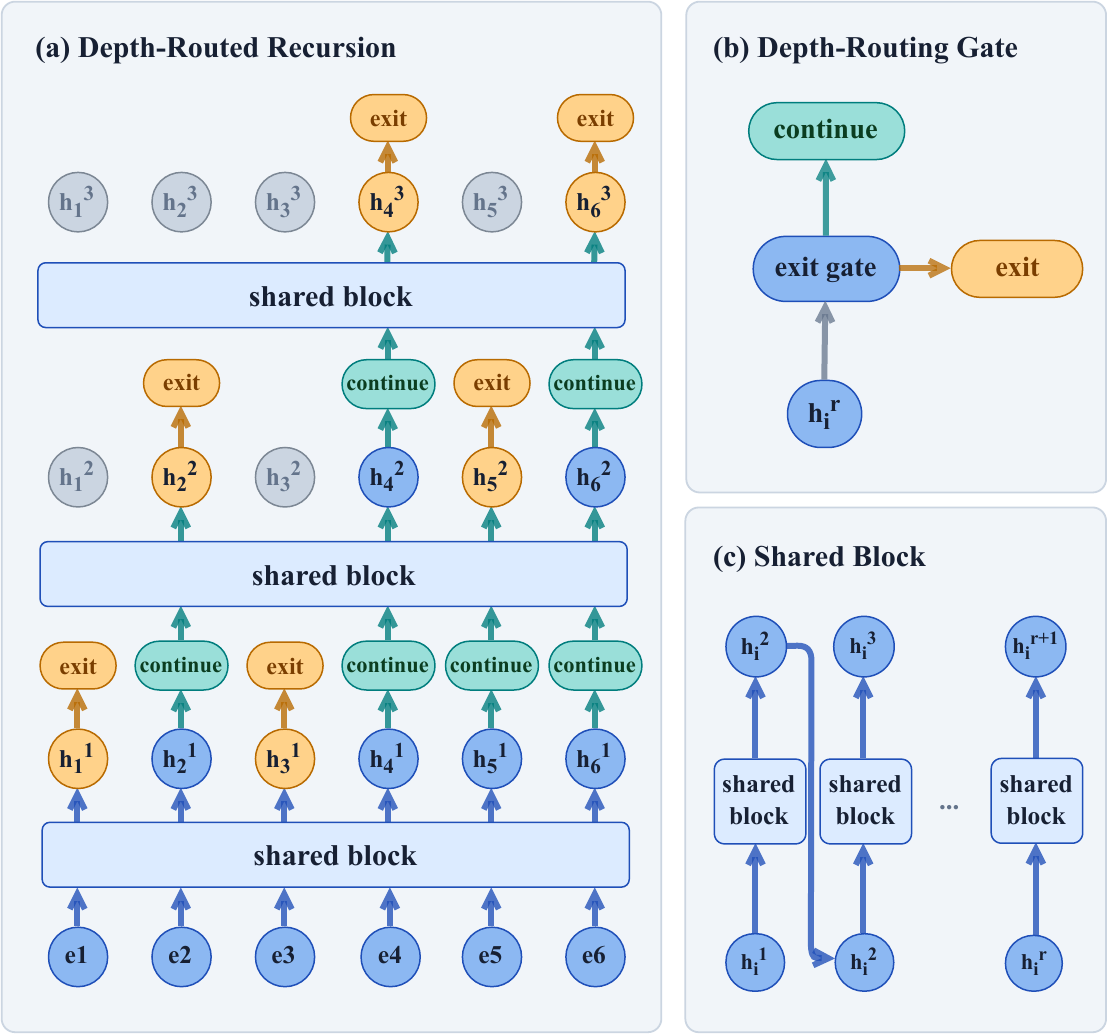}
  \caption{Overview of LoopRec. At each recursion boundary, depth-routing gates decide which positions stop and which continue through the shared block.}
  \label{fig:LoopRec-overview}
\end{figure}

To realize this adaptive-depth mechanism within a single model, LoopRec treats the backbone block as a reusable computation unit and shares it across recursions. Let $\mathcal{B}(\cdot)$ denote a shared backbone block with $L$ layers, and let $R$ be the maximum number of recursions. We denote this layer--recursion configuration as \texttt{LxRy}. For example, \texttt{L2R3} uses a two-layer shared block for up to three recursions. Given $\mathbf{X}$, LoopRec proceeds through three components. \textbf{Initial Context Encoding} first applies $\mathcal{B}(\cdot)$ to the full sequence to obtain contextualized states for all positions. \textbf{Depth-Routed Recursion} then applies adaptive depth routing at each recursion boundary: the depth router makes a stop-or-continue decision for each active position and re-applies the shared block only to positions that continue, until all positions stop or the recursion budget is reached. Finally, \textbf{Depth-Specific Readout} forms the representation used for next-item prediction from the state at the selected depth. A position that stops after recursion $r$ has effectively received $rL$ layers of sequential encoding. Figure~\ref{fig:LoopRec-overview} provides an overview.

\textbf{Initial Context Encoding.}
LoopRec begins with a shared shallow pass over the full sequence:
\begin{equation*}
\mathbf{H}^{(1)}=\mathcal{B}(\mathbf{X}),
\end{equation*}
where
$\mathbf{H}^{(1)}=[\mathbf{h}^{(1)}_{1},\dots,\mathbf{h}^{(1)}_{T}]$.
This initial pass provides the common contextualized states on which
subsequent routing decisions for individual history items are based.

\textbf{Depth-Routed Recursion.}
Starting from the contextualized states $\mathbf{H}^{(1)}$, LoopRec performs depth-routed recursion for up to $R-1$ additional steps. At recursion boundary $r\in\{1,\dots,R-1\}$, the depth router examines the current state of each active position and decides whether that position should stop at the current depth or continue to a deeper recursion. For position $i$, let $\mathbf{h}_{i}^{(r)}$ denote its state after recursion $r$. The depth router computes a routing score
\begin{equation*}
s^{(r)}_{i}=g(\mathbf{h}^{(r)}_{i}+\boldsymbol{\rho}_{r}),
\end{equation*}
where $g(\cdot)$ is the router head and $\boldsymbol{\rho}_{r}$ is a learnable embedding of the recursion index. This score is converted into an exit probability
\begin{equation*}
q^{(r)}_{i,\mathrm{exit}} = 1-\sigma(s^{(r)}_{i}),
\end{equation*}
where $\sigma(\cdot)$ is the logistic sigmoid. A position stops at boundary $r$ if $q^{(r)}_{i,\mathrm{exit}}\ge 0.5$. Otherwise, it is sent to the next recursion.

Let $\mathcal{A}^{(r+1)}$ denote the positions that continue after boundary $r$. Then
\begin{equation*}
\mathcal{A}^{(r+1)}
=\{i\in\mathcal{A}^{(r)}: q^{(r)}_{i,\mathrm{exit}}<0.5\}.
\end{equation*}
Only these continuing positions are updated by re-applying the shared block to their states from the previous recursion:
\begin{equation*}
\mathbf{H}^{(r+1)}_{\mathcal{A}^{(r+1)}}
=\mathcal{B}\!\left(\mathbf{H}^{(r)}_{\mathcal{A}^{(r+1)}}\right).
\end{equation*}
Thus, each recursion takes as input the states from the previous recursion, but only for the positions that continue. The selected subsequence preserves its original order under the causal mask, while stopped positions keep their current states and are not updated further. Consequently, $\mathcal{A}^{(1)}\supseteq\mathcal{A}^{(2)}\supseteq\dots \supseteq\mathcal{A}^{(R)}$: the set of continuing positions shrinks as recursion proceeds. At the final recursion, all remaining positions are forced to stop.

\textbf{Depth-Specific Readout.}
After recursion terminates, each position is associated with a selected depth $d_i\in\{1,\dots,R\}$. LoopRec then forms the final representation for position $i$ by combining its input token with the
state at the selected depth:
\begin{equation*}
\bar{\mathbf{z}}_{i}
=\mathbf{x}_{i}+\omega_i\mathbf{h}^{(d_i)}_{i},
\end{equation*}
where $\omega_i$ is a depth-dependent readout weight.
This readout uses the state selected by depth-routed recursion while retaining a residual connection to the input token.
Although the forward pass selects a hard depth $d_i$ for each position, the readout weight $\omega_i$ maintains differentiability with respect to the router probabilities to enable gradient-based training. Its detailed definition is given in Appendix~\ref{app:routing-readout}.
The recursive encoding and readout procedure is summarized in Algorithm~\ref{alg:looprec}.

\subsection{Training Objective}
\label{sec:objective}

The key additional component in LoopRec, compared with a fixed-depth sequential recommender, is the depth router. Training therefore needs to ensure that this router is both useful and learnable. First, it should allocate different positions to different depths. Otherwise LoopRec collapses toward nearly fixed-depth behavior and loses
the item-level adaptivity motivated in Sec.~\ref{sec:rethinking}. Second, it should remain trainable throughout optimization. Because routing decisions are made recursively, gates that become too hard too early can lock the model into unstable depth assignments before the hidden states are well formed. The objective design below is organized around these two requirements.

\textbf{Avoiding Depth Collapse.}
To avoid depth collapse, we regularize depth usage along the realized route. The idea is to assign each position a soft depth mass over the depths it actually visits, and then match the aggregate soft usage to the empirical hard assignment frequencies across positions. We first define the continuation mass that reaches each boundary.

Let $d_i\in\{1,\ldots,R\}$ denote the selected depth of position
$i$. We first define the remaining continuation mass before each evaluated boundary.
Let
\begin{equation*}
c_i^{(1)}=1,
\qquad
c_i^{(r+1)}=c_i^{(r)}(1-q^{(r)}_{i,\mathrm{exit}})
\ \text{for}\ r<d_i.
\end{equation*}
Thus $c_i^{(r)}$ is the amount of route mass that reaches boundary $r$ without exiting earlier.

Using this continuation mass, we define the soft depth mass assigned to position $i$ at depth $r$ as
\begin{equation*}
\alpha_{i,r}=
\begin{cases}
c_i^{(r)}q^{(r)}_{i,\mathrm{exit}}, & r<d_i,\\
c_i^{(d_i)}, & r=d_i,\\
0, & r>d_i.
\end{cases}
\end{equation*}
Thus $\alpha_{i,r}$ records the route mass assigned to depth $r$ using only the depths that are actually evaluated.

Let $\mathcal{I}$ be the set of valid positions and let $N=|\mathcal{I}|$. We then define the average soft depth usage at depth $r$ as
\begin{equation*}
P_r=\frac{1}{N}\sum_{i\in\mathcal{I}}\alpha_{i,r},
\end{equation*}
where $P_r$ is the batch-averaged soft usage of depth $r$: it aggregates
the realized-route mass $\alpha_{i,r}$ over valid positions and captures
the average soft route mass assigned to that depth.

The corresponding hard assignment frequency is
\begin{equation*}
f_r=\frac{1}{N}\sum_{i\in\mathcal{I}}\mathbb{I}[d_i=r].
\end{equation*}
Here $f_r$ is the fraction of valid positions whose selected depth is
$r$, i.e., the empirical frequency of hard routing assignments to depth
$r$ across the current batch.

The load-balancing regularizer~\cite{shazeer2017outrageously,fedus2022switch} is
\begin{equation*}
\mathcal{L}_{\mathrm{bal}}=R\sum_{r=1}^{R} P_r f_r.
\end{equation*}
Here $P_r$ measures the average soft mass assigned to depth $r$, while
$f_r$ measures how often depth $r$ is selected by the hard route.
Minimizing $\mathcal{L}_{\mathrm{bal}}$ discourages the router from
concentrating depth usage on a narrow range of depths.

\textbf{Avoiding Premature Gate Saturation.}
A learnable router should not become overly confident before the hidden states are stable. To discourage this behavior, we penalize large binary-gate log partitions, following the router z-loss used to stabilize sparse routed models~\cite{zoph2022stmoe} and later adopted in MoR~\cite{bae2025mor}. Using the routing score $s_i^{(r)}=g(\mathbf{h}_i^{(r)}+\boldsymbol{\rho}_r)$ introduced in the depth-routing step, the resulting anchored binary gate has log partition $\operatorname{softplus}(s_i^{(r)})=\log(1+\exp(s_i^{(r)}))$. Let
$\mathcal{R}_{+}=\{r:1\le r<R,\ |\mathcal{A}^{(r)}|>0\}$. Then
\begin{equation}
\mathcal{L}_{z}
=\frac{1}{|\mathcal{R}_{+}|}
\sum_{r\in\mathcal{R}_{+}}
\frac{1}{|\mathcal{A}^{(r)}|}
\sum_{i\in\mathcal{A}^{(r)}}
\operatorname{softplus}(s_i^{(r)})^2 .
\end{equation}
This term keeps routing decisions from becoming too hard too early in
training.

\begin{table}[t]
\centering
\small
\caption{Statistics of experimental datasets.}
\label{tab:dataset-stats}
\begin{tabular*}{\columnwidth}{@{\hspace{\tabcolsep}\extracolsep{\fill}}lrrrrr@{\hspace{\tabcolsep}}}
\toprule
\textbf{Dataset} & \textbf{Users} & \textbf{Items}
& \textbf{Interactions} & \textbf{Avg.\ Len.} & \textbf{Sparsity} \\
\midrule
Software & 13,444 & 9,706 & 227,637 & 16.93 & 99.83\% \\
Office & 11,468 & 28,279 & 190,465 & 16.61 & 99.94\% \\
Patio & 14,454 & 37,179 & 229,086 & 15.85 & 99.96\% \\
\bottomrule
\end{tabular*}
\end{table}

\textbf{Overall Objective.}
Combining the recommendation target with the two router-side regularizers, LoopRec optimizes
\begin{equation}
\mathcal{L}
=\mathcal{L}_{\mathrm{rec}}
+\lambda_{\mathrm{bal}}\mathcal{L}_{\mathrm{bal}}
+\lambda_{z}\mathcal{L}_{z},
\end{equation}
where $\lambda_{\mathrm{bal}}$ and $\lambda_{z}$ control the strengths of the depth-diversity and gate-stabilization regularizers, respectively. Here $\mathcal{L}_{\mathrm{rec}}$ remains the only item-prediction supervision, while the additional terms regularize only the routing process.

\textbf{Routing Warm-Up.}
In addition to the objective above, we use a simple warm-up strategy to avoid unstable routing at the beginning of training. Let $W_{\mathrm{router}}$ denote the warm-up length. For the first $W_{\mathrm{router}}$ epochs, all valid positions are forced to the maximum recursion depth, routing decisions are bypassed, and the
router-side regularizers are disabled. Training in this phase therefore
reduces to next-item prediction under a fixed maximum-depth recursive
pass. After warm-up, routing is enabled and the full objective above is
optimized. We treat $W_{\mathrm{router}}$ as a training
hyperparameter and select it with validation NDCG@20.

\section{Experiments}

\subsection{Experimental Setup}

\mypara{Datasets}
We evaluate LoopRec on three Amazon Reviews 2023 categories~\cite{hou2024bridging}: Software, Office Products (Office), and Patio, Lawn \& Garden (Patio). These categories provide varied sequence and sparsity regimes under a shared data source and preprocessing protocol.
For each category, we sort interactions by time and retain users with at least $10$ interactions. We use a leave-one-out split for validation and test sets.
 During evaluation, we trim each history to warm-start items so that all conditioning items have appeared in the training split. The maximum sequence length is set to $50$.
Dataset statistics are summarized in Table~\ref{tab:dataset-stats}.

\mypara{Evaluation Metrics}
We evaluate all models with full-ranking top-$K$ metrics~\cite{krichene2020sampled}. MRR measures
the rank of the first relevant item, while NDCG accounts for the
position-discounted ranking quality of the held-out next item~\cite{jarvelin2002cumulated}. We
report MRR@10, MRR@20, NDCG@10, and NDCG@20 in the main results.
Model selection and early stopping use validation NDCG@20.

\mypara{Backbones}
LoopRec is instantiated on four widely used backbone families to test its compatibility with different sequence modeling architectures:
\begin{itemize}[leftmargin=1em, labelsep=0.4em, itemindent=0pt]
\item \textbf{SASRec}~\cite{kang2018sasrec} is a self-attention sequential recommender and serves as the main attention anchor. It gives the most direct comparison with a Transformer recommender of fixed depth.
\item \textbf{BSARec}~\cite{shin2024bsarec} is a frequency-aware self-attention variant. It adds a frequency-domain bias to the SASRec family and tests transfer within attention models.
\item \textbf{NextItNet}~\cite{yuan2019nextitnet} is a causal convolutional recommender. It uses causal convolutions and represents the convolutional family.
\item \textbf{LRURec}~\cite{yue2024linear} is a linear recurrent recommender. Together with NextItNet, it tests whether LoopRec remains applicable beyond attention backbones.
\end{itemize}
\mypara{Implementation Details}
Training is capped at 100 epochs with early stopping patience 10. Test metrics are reported from the checkpoint selected by the validation
criterion above. Backbone-specific wrapping choices and detailed hyperparameter settings are provided in Appendix~\ref{app:backbone-details}.

\begin{table*}[t]
\centering
\small
\setlength{\tabcolsep}{0pt}
\renewcommand{\arraystretch}{1.0}
\caption{Main results across three Amazon benchmarks, comparing each fixed-depth backbone with its LoopRec counterpart. \textbf{M@K} and \textbf{N@K} denote MRR@K and NDCG@K, and values are test means over five runs. The ``*'' denotes the statistical significance ($p < 0.05$) of the results of LoopRec compared to the corresponding fixed-depth backbone under a paired t-test. Within each backbone block, the better value is shown in \textbf{bold}.}
\label{tab:main-results}
\begin{tabular*}{0.98\textwidth}{@{\extracolsep{\fill}}llrrrrrrrrrrrr@{}}
\toprule
\multirow{2}{*}{\textbf{Backbone}} & \multirow{2}{*}{\textbf{Model}}
& \multicolumn{4}{c}{\textbf{Software}}
& \multicolumn{4}{c}{\textbf{Office}}
& \multicolumn{4}{c}{\textbf{Patio}} \\
\cmidrule(lr){3-6}\cmidrule(lr){7-10}\cmidrule(l){11-14}
& & \textbf{M@10} & \textbf{M@20} & \textbf{N@10} & \textbf{N@20}
  & \textbf{M@10} & \textbf{M@20} & \textbf{N@10} & \textbf{N@20}
  & \textbf{M@10} & \textbf{M@20} & \textbf{N@10} & \textbf{N@20} \\
\midrule
\multirow{3}{*}{SASRec}
& \textsc{Vanilla}
& 0.0589 & 0.0629 & 0.0796 & 0.0944
& 0.0508 & 0.0513 & 0.0562 & 0.0582
& 0.0394 & 0.0399 & 0.0429 & 0.0448 \\
& +LoopRec
& \textbf{0.0613} & \textbf{0.0659} & \textbf{0.0830} & \textbf{0.0998}
& \textbf{0.0523} & \textbf{0.0530} & \textbf{0.0571} & \textbf{0.0595}
& \textbf{0.0411} & \textbf{0.0416} & \textbf{0.0441} & \textbf{0.0458} \\
& Improv.
& +4.14\%* & +4.77\%* & +4.27\%* & +5.68\%*
& +3.03\%* & +3.24\%* & +1.49\%* & +2.20\%*
& +4.42\%* & +4.26\%* & +2.75\%* & +2.19\%* \\
\midrule
\multirow{3}{*}{BSARec}
& \textsc{Vanilla}
& 0.0607 & 0.0657 & 0.0803 & 0.0987
& 0.0508 & 0.0516 & 0.0565 & 0.0596
& 0.0391 & 0.0398 & 0.0426 & 0.0453 \\
& +LoopRec
& \textbf{0.0619} & \textbf{0.0665} & \textbf{0.0828} & \textbf{0.0998}
& \textbf{0.0519} & \textbf{0.0527} & \textbf{0.0571} & \textbf{0.0602}
& \textbf{0.0415} & \textbf{0.0422} & \textbf{0.0445} & \textbf{0.0469} \\
& Improv.
& +1.94\%* & +1.22\%* & +3.11\%* & +1.13\%*
& +2.13\%* & +2.21\%* & +0.92\% & +0.97\%
& +6.30\%* & +6.03\%* & +4.46\%* & +3.48\%* \\
\midrule
\multirow{3}{*}{NextItNet}
& \textsc{Vanilla}
& 0.0431 & 0.0479 & 0.0608 & 0.0782
& 0.0225 & 0.0235 & 0.0274 & 0.0313
& 0.0129 & 0.0136 & 0.0166 & 0.0192 \\
& +LoopRec
& \textbf{0.0460} & \textbf{0.0506} & \textbf{0.0648} & \textbf{0.0817}
& \textbf{0.0256} & \textbf{0.0267} & \textbf{0.0312} & \textbf{0.0354}
& \textbf{0.0193} & \textbf{0.0201} & \textbf{0.0231} & \textbf{0.0263} \\
& Improv.
& +6.77\%* & +5.59\%* & +6.51\%* & +4.37\%*
& +13.88\%* & +13.59\%* & +14.04\%* & +13.18\%*
& +49.07\%* & +47.94\%* & +39.06\%* & +37.02\%* \\
\midrule
\multirow{3}{*}{LRURec}
& \textsc{Vanilla}
& 0.0575 & 0.0625 & 0.0777 & 0.0958
& 0.0502 & 0.0508 & 0.0550 & 0.0573
& 0.0407 & 0.0412 & 0.0442 & 0.0461 \\
& +LoopRec
& \textbf{0.0620} & \textbf{0.0670} & \textbf{0.0814} & \textbf{0.1001}
& \textbf{0.0509} & \textbf{0.0517} & \textbf{0.0558} & \textbf{0.0584}
& \textbf{0.0409} & \textbf{0.0415} & \textbf{0.0449} & \textbf{0.0470} \\
& Improv.
& +7.68\%* & +7.20\%* & +4.81\%* & +4.51\%*
& +1.43\%* & +1.73\%* & +1.42\%* & +1.85\%*
& +0.69\% & +0.83\% & +1.58\%* & +1.87\%* \\
\bottomrule
\end{tabular*}
\end{table*}

\subsection{Main Results}

Table~\ref{tab:main-results} compares four fixed-depth backbones with their LoopRec counterparts across three Amazon datasets. We highlight three main observations below.

\noindent\textbf{(1) LoopRec consistently improves fixed-depth baselines.} In all $48$ backbone--dataset--metric comparisons, LoopRec yields positive improvements, and $44$ of them are statistically significant. The
relative gains range from $0.69\%$ to $49.07\%$, showing that the benefit is not limited to a particular backbone family or dataset. The gains are also aligned across MRR and NDCG within each backbone block, indicating that the improvement is not an artifact of a single metric. Overall, these results show that replacing uniform fixed-depth computation with item-level adaptive depth leads to consistently stronger next-item ranking performance.

\noindent\textbf{(2) Gains transfer across backbone architectures.}
For attention-based models, LoopRec improves NDCG@20 by
$2.19$--$5.68\%$ on SASRec and $0.97$--$3.48\%$ on BSARec.
The gains on BSARec show that adaptive computation remains useful
alongside its frequency-domain inductive bias.
Beyond attention, LRURec gains $1.85$--$4.51\%$ on NDCG@20,
demonstrating compatibility with linear recurrent sequence modeling.
NextItNet shows particularly large relative gains on Office and Patio,
reaching $13.18\%$ and $37.02\%$, respectively.
On Patio, its NDCG@20 increases from $0.0192$ to $0.0263$.
Together, these results support the applicability of item-level
adaptive recursion across backbone families, whether they use
self-attention, linear recurrence, or causal convolution to
integrate sequence context when predicting the next item.

\noindent\textbf{(3) The largest gains appear on the sparsest dataset.}
The improvement magnitude varies across datasets. The gains are moderate on Software and Office for most strong baselines, but become more pronounced on Patio, which has the highest sparsity, the largest item vocabulary, and the shortest average history length in Table~\ref{tab:dataset-stats}. Averaged over backbones, the NDCG@20 improvement increases from $3.92\%$ on Software and $4.55\%$ on Office to $11.14\%$ on Patio. This suggests that adaptive-depth computation is more useful when user histories provide fewer interaction signals. In such data, different history items may carry uneven amounts of useful context, so applying the same computation depth to every position can be less suitable. LoopRec can instead stop processing simple positions at an earlier recursion boundary and spend deeper computation on positions that need more contextual modeling.

\subsection{What Shapes LoopRec's Learned Depth?}
\label{sec:analysis-factors}

Having established that LoopRec improves ranking accuracy, we next ask what its learned depth allocation captures. Our analysis focuses on position- and item-level factors, and finally connects LoopRec's learned routing back to the saturation-depth probe of Section~\ref{sec:rethinking}. Unless otherwise specified, the analyses in this section use SASRec as the backbone and Office as the dataset.

The router uses multiple effective depths rather than a single near-fixed path (Appendix~\ref{app:depth-distribution}). We examine this variation from three views:
the relative position of the item in the history, its training-time
frequency, and the depth-sensitivity pattern that the same item shows
under a fixed-depth reference recommender.

\subsubsection{Depth by Sequence Position}
\label{sec:analysis-position-depth}

A natural starting point is the relative position of an item in the
history. For each test sequence, we divide the valid history into ten
deciles, from the earliest interactions to the most recent ones, and
compute the average selected recursion round within each decile. To
give a broader view than a single setting, we report four
SASRec-backbone cases covering Office and Patio under two \texttt{LxRy}
configurations, where each recursion round applies $x$ shared blocks and at most $y$ rounds are available.

\begin{figure}[t]
  \centering
  \includegraphics[width=\linewidth]{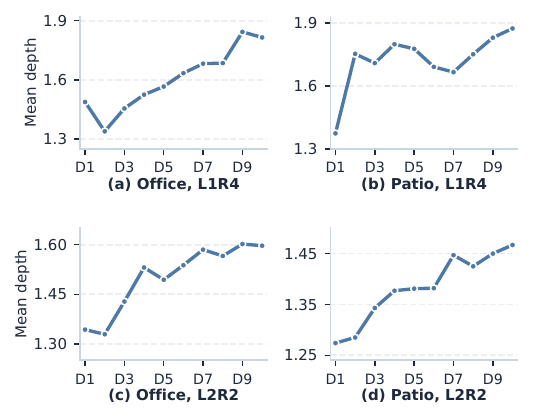}
  \caption{Average selected recursion round across relative-position deciles (D1: earliest, D10: most recent), on Office and Patio under two \texttt{LxRy} configurations.}
  \label{fig:sasrec-position-wise-depth}
\end{figure}

\noindent\textbf{Recent history items are routed deeper on average.}
Figure~\ref{fig:sasrec-position-wise-depth} shows an average tendency
for later history deciles to receive larger selected rounds in the four
cases. This suggests that relative position is associated with depth
allocation: recent interactions tend to carry stronger signals about
the user's current preference, so LoopRec assigns more of its depth
budget to the later part of the history. The relation is not a
per-item rule, but it shows that LoopRec does not allocate depth
uniformly across the whole history.

\subsubsection{Depth by Item-Level Factors}
\label{sec:analysis-item-signals}

Position alone does not fully describe a history item. Two items at
the same relative position can differ in how often they appear during
training and in how much context is needed to interpret them. We
therefore ask whether routed depth also correlates with item-level
signals. As a focused diagnostic, we bucket test-set history tokens
by the corresponding item's training-time interaction frequency and
report the resulting depth statistics in
Table~\ref{tab:item-popularity-depth}.

\noindent\textbf{Frequent items are assigned shallower depths, while rarer items receive
more computation.} As training frequency increases, the mean selected
depth generally decreases from $2.160$ in the single-occurrence bucket
to $1.330$ in the $100\!+\!$ bucket. The share of depth-1 assignments rises
from $50.3\%$ to $88.6\%$, while the share of depth-4 assignments drops from
$29.1\%$ to $10.7\%$. Routed depth therefore correlates not only with
position but also with the sparsity of item-level interaction signals.
This suggests that frequent items form stable representations
with little additional computation, while rarer items more often
benefit from later recursive updates.
Appendix~\ref{app:item-routing-case} complements these aggregate
statistics with a category-level view and a full sequence example
illustrating the routing patterns.

The observed relationships between depth, recency, and frequency
motivate testing whether simple routing rules based on these signals
can achieve comparable recommendation quality.
On Office with SASRec, LoopRec achieves higher mean NDCG@20 than
either the position-based or frequency-based heuristic over five seeds.
The experimental setup and full results are provided in
Appendix~\ref{app:heuristic-routing}.

\begin{table}[t]
\centering
\small
\caption{Selected depth by item training-time frequency on Office. Depth columns report the share of tokens assigned to each depth.}
\label{tab:item-popularity-depth}
\begin{tabular*}{\columnwidth}{@{\hspace{\tabcolsep}\extracolsep{\fill}}lrrrr@{\hspace{\tabcolsep}}}
\toprule
\textbf{Train freq.} & \textbf{Tokens} & \textbf{Mean depth}
& \textbf{Depth 1} & \textbf{Depth 4} \\
\midrule
1 & 505 & 2.160 & 50.3\% & 29.1\% \\
2--4 & 1,202 & 2.084 & 53.7\% & 27.0\% \\
5--9 & 1,603 & 2.135 & 53.3\% & 29.3\% \\
10--19 & 2,247 & 2.008 & 60.8\% & 28.1\% \\
20--49 & 4,123 & 1.818 & 69.2\% & 24.1\% \\
50--99 & 3,668 & 1.577 & 79.1\% & 17.7\% \\
100+ & 11,659 & 1.330 & 88.6\% & 10.7\% \\
\bottomrule
\end{tabular*}
\end{table}

\subsubsection{Alignment with Saturation Depth}
\label{sec:analysis-saturation-callback}

The previous analyses show that LoopRec assigns deeper routes to recent
and to rarer items. We now return to the saturation-depth probe in
Section~\ref{sec:rethinking} and ask whether the same two factors
emerge in the fixed-depth reference's own depth-sensitivity pattern.
Such agreement would indicate that LoopRec's learned routing is aligned
with the depth demand that the fixed-depth recommender reveals only after
training. Recall that
$\ell_i^\star$ is the earliest layer after which freezing item $i$'s
updates preserves the full-depth recommendation set. A larger
$\ell_i^\star$ marks an item that stabilizes later inside the
fixed-depth reference.

\begin{figure}[t]
  \centering
  \includegraphics[width=\linewidth]{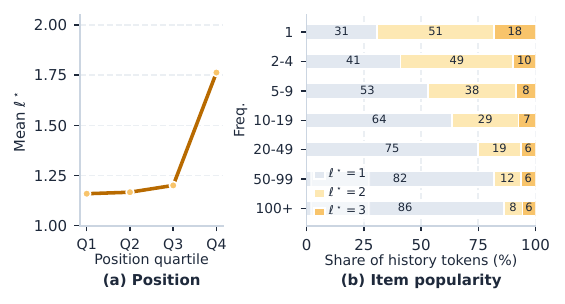}
  \caption{Saturation depth on Office under a three-layer SASRec. (a)~Mean depth across relative-position quartiles. (b)~Depth shares across training-frequency buckets.}
  \label{fig:saturation-factor-callback}
\end{figure}

\noindent\textbf{The position effect also appears in saturation
depth.} Grouping history tokens into four relative-position quartiles
(Figure~\ref{fig:saturation-factor-callback}(a)), the first three
quartiles have low mean saturation depths of $1.16$, $1.17$, and
$1.20$, while the most recent quartile rises to $1.76$. The
$\ell^\star=3$ share is also concentrated in the recent quartile
($30.3\%$), with essentially no full-depth saturation in the earlier
three. Positions near the target are therefore more likely to require
deeper processing under the fixed-depth reference, in the same direction
as LoopRec's routed-depth trend in
Section~\ref{sec:analysis-position-depth}.

\noindent\textbf{The popularity effect is also visible under the same
diagnostic.} Using the same training-frequency buckets as
Table~\ref{tab:item-popularity-depth},
Figure~\ref{fig:saturation-factor-callback}(b) shows that rarer items
have larger saturation depths and larger full-depth shares. In the
single-occurrence bucket, the mean saturation depth is $1.873$ with an
$\ell^\star=3$ share of $18.0\%$. In the $100\!+\!$ bucket, the mean
drops to $1.197$, the $\ell^\star=3$ share to $5.7\%$, and the
$\ell^\star=1$ share reaches $86.1\%$. This mirrors the popularity
trend in Section~\ref{sec:analysis-item-signals}: rare items receive
deeper routes from LoopRec and also stabilize later in the
saturation-depth probe.

\noindent\textbf{LoopRec's learned routing aligns with the
saturation-depth probe.} Taken together, the router allocates more
computation to recent and rare items, and the same groups also show
larger saturation depths under the fixed-depth reference. This
agreement supports a \emph{demand-aligned} reading of LoopRec's depth
allocation: extra computation is directed to the items that the
saturation-depth probe also marks as sensitive to additional processing.

\subsection{Further Analysis}
\label{sec:further-analysis}

Having characterized LoopRec's learned depth allocation, we now turn to ablation studies and computational cost. We examine the contributions of shared recursion, item-level depth allocation, routing warm-up, and router-side regularizers, and then report the computational savings and measured encoder latency.

\subsubsection{Ablation Studies}
\label{sec:analysis-ablation}

We first compare sequence-level looping with LoopRec's item-level depth allocation. We then ablate
the routing warm-up that bridges full-depth training to learned depth decisions
and the two router-side regularizers (load balancing and z-loss). All studies
in this subsection use the SASRec-based recipe.

\begin{table}[h]
\centering
\small
\caption{Comparison of sequence-level looping and item-level depth allocation. Vanilla denotes the original SASRec. Results are five-seed test means. M and N denote MRR and NDCG, respectively.}
\label{tab:ablation-shared-recursion}
\setlength{\tabcolsep}{2pt}
\begin{tabularx}{\columnwidth}{@{}l*{6}{>{\raggedleft\arraybackslash}X}@{}}
\toprule
\multirow{2}{*}{\textbf{Model}}
& \multicolumn{2}{c}{\textbf{Software}}
& \multicolumn{2}{c}{\textbf{Office}}
& \multicolumn{2}{c}{\textbf{Patio}} \\
\cmidrule(lr){2-3}\cmidrule(lr){4-5}\cmidrule(lr){6-7}
& \textbf{M@20} & \textbf{N@20}
& \textbf{M@20} & \textbf{N@20}
& \textbf{M@20} & \textbf{N@20} \\
\midrule
\tablegroupcolor LoopRec & \textbf{0.0659} & \textbf{0.0998}
& \textbf{0.0530} & \textbf{0.0595}
& \textbf{0.0416} & \textbf{0.0458} \\
Seq-level Loop & 0.0634 & 0.0950
& 0.0525 & 0.0594
& 0.0390 & 0.0447 \\
\textsc{Vanilla} & 0.0629 & 0.0944
& 0.0513 & 0.0582
& 0.0399 & 0.0448 \\
\bottomrule
\end{tabularx}
\end{table}

\mypara{Sequence-level and item-level allocation}
LoopRec combines weight-sharing through block reuse with item-level depth allocation. We compare it with a sequence-level looped Transformer baseline (\textit{Seq-level Loop}), which reuses an $L$-layer SASRec block for $R$ full-sequence passes, with no router or item-level early exit. Training and evaluation follow the main SASRec protocol. We search over the same \texttt{LxRy} configuration set as LoopRec (Appendix~\ref{app:backbone-details}), selecting by validation NDCG@20.
Table~\ref{tab:ablation-shared-recursion} shows that \textit{Seq-level Loop} improves both MRR@20 and NDCG@20 over Vanilla on Software and Office, but yields no improvement on Patio. Full LoopRec achieves the highest mean for both metrics on all three datasets. Compared with \textit{Seq-level Loop}, LoopRec gains $0.0025$ in MRR@20 and $0.0048$ in NDCG@20 on Software. The gains on Office are small, while on Patio MRR@20 increases from $0.0390$ to $0.0416$. These results suggest that weight sharing through recursion accounts for part of the improvement, and item-level depth allocation adds further improvements that vary in magnitude across the three datasets.

\begin{table}[h]
\centering
\small
\caption{Ablation of routing warm-up and router-side regularizers. Results are five-seed test means. M and N denote MRR and NDCG, respectively.}
\label{tab:ablation-routing-components}
\setlength{\tabcolsep}{2pt}
\begin{tabularx}{\columnwidth}{@{}l*{6}{>{\raggedleft\arraybackslash}X}@{}}
\toprule
\multirow{2}{*}{\textbf{Ablation}}
& \multicolumn{2}{c}{\textbf{Software}}
& \multicolumn{2}{c}{\textbf{Office}}
& \multicolumn{2}{c}{\textbf{Patio}} \\
\cmidrule(lr){2-3}\cmidrule(lr){4-5}\cmidrule(lr){6-7}
& \textbf{M@20} & \textbf{N@20}
& \textbf{M@20} & \textbf{N@20}
& \textbf{M@20} & \textbf{N@20} \\
\midrule
\tablegroupcolor LoopRec & \textbf{0.0659} & \textbf{0.0998}
& \textbf{0.0530} & \textbf{0.0595}
& \textbf{0.0416} & \textbf{0.0458} \\
w/o warm-up & 0.0615 & 0.0942
& 0.0520 & 0.0583
& 0.0401 & 0.0450 \\
w/o $\mathcal{L}_{\mathrm{bal}}$ & 0.0655 & 0.0995
& 0.0507 & 0.0576
& 0.0385 & 0.0431 \\
w/o $\mathcal{L}_{z}$ & 0.0639 & 0.0973
& 0.0507 & 0.0574
& 0.0387 & 0.0432 \\
w/o $\mathcal{L}_{\mathrm{bal}}$ \& $\mathcal{L}_{z}$ & 0.0632 & 0.0963
& 0.0505 & 0.0574
& 0.0386 & 0.0431 \\
\bottomrule
\end{tabularx}
\end{table}

\mypara{Routing warm-up and router-side regularizers}
Table~\ref{tab:ablation-routing-components} examines routing warm-up, load balancing $\mathcal{L}_{\mathrm{bal}}$, and router z-loss $\mathcal{L}_{z}$. Removing warm-up lowers both metrics on all three datasets, dropping MRR@20 by $4.03\%$ and NDCG@20 by $3.11\%$ on average. Software shows the largest decrease in both metrics. Removing either regularizer lowers both metrics on all three datasets, with larger relative drops on Office and Patio than on Software. Removing both regularizers reduces MRR@20 by $5.36\%$ and NDCG@20 by $4.25\%$ on average across datasets.

\subsubsection{Computation and Latency}

Across the three measured settings, LoopRec reduces selected compute to $0.407$--$0.755\times$ the fixed-depth reference, while encoder latency spans $0.869$--$0.995\times$. Compute savings therefore translate only partly into wall-clock gains. Complete measurements are reported in Appendix~\ref{app:runtime-latency}.

\section{Related Work}

\mypara{Sequential Recommendation}
Sequential recommendation models an ordered interaction history to
rank the next item. Early factorized and neural approaches capture
local transition structure via Markov chains, recurrent networks, or
convolutional encoders~\cite{rendle2010fpmc,hidasi2016gru4rec,tang2018caser,yuan2019nextitnet,li2017neural,liu2018stamp}.
Transformer-based recommenders such as SASRec and BERT4Rec then
strengthen long-range dependency modeling with self
attention~\cite{kang2018sasrec,sun2019bert4rec,li2020time,zhou2020s3,xie2022contrastive,qiu2022contrastive}. More recent
fixed-depth backbones refine the encoder design with explicit
inductive biases, for instance through frequency-domain or filtering
mechanisms~\cite{shin2024bsarec,zhou2022filter,du2023frequency}. These methods focus on \emph{how} to build stronger sequence encoders under a fixed computation pattern
: once the architecture is chosen, each history
position is processed by the same depth budget before scoring the next
item. LoopRec addresses an orthogonal question. Keeping the
recommendation objective and backbone family intact, it asks whether
a sequential encoder can allocate recursive computation at the
\emph{item} level, so that different positions in the same history
need not receive the same depth.

\mypara{Looped and Adaptive-Depth Transformers}
Looped Transformers study the repeated application of shared or
partially shared Transformer blocks.
The Universal Transformer combines shared recurrent transformations with self-attention and supports adaptive per-position halting~\cite{dehghani2019universal}, building on adaptive computation time~\cite{graves2016adaptive}. Later work shows that
looped Transformers can implement non-trivial algorithmic
computation~\cite{giannou2023looped,yang2024looped}. Recent looped Transformer and
looped language-model studies connect repeated depth with latent
computation and reasoning~\cite{saunshi2025latent,zhu2025scaling}. In parallel,
adaptive-depth Transformers explore early exiting or routed computation
allocation~\cite{elbayad2020depth,raposo2024mixture,schuster2022confident,xin2020deebert,liu2020fastbert,zhou2020bert}. Mixture-of-Recursions (MoR) combines shared recursion with token-level adaptive computation for language modeling~\cite{bae2025mor}. Its token-choice router assigns a total recursion depth upfront, while its expert-choice router selects continuing tokens at each step under prescribed capacities. LoopRec uses a shared binary stopping gate conditioned on each item's current contextualized state and recursion index, allowing decisions to be updated across recursions without per-step top-$k$ selection. We investigate this design for computation allocation over recommendation histories, supported by item-level diagnostics and evaluation across multiple backbone families.

\section{Conclusion}

We revisited fixed-depth computation in sequential recommendation
through diagnostic experiments that reveal heterogeneous depth sensitivity
across history items. Motivated by these findings, we proposed \textbf{LoopRec},
which uses weight-shared recursion with learned item-wise routing to
allocate computation adaptively. Across three Amazon benchmarks and
four backbone families, LoopRec consistently improves ranking performance
over fixed-depth counterparts. Its learned routes allocate greater depth
to recent and less frequent items, consistent with the patterns observed
in these diagnostics.
LoopRec also reduces the encoder computation budget in the measured
configurations, although latency gains vary. Future work includes
developing sparsity-aware kernels and learning the maximum recursion budget,
which is currently fixed per configuration before model training begins.

\bibliographystyle{ACM-Reference-Format}
\bibliography{bib/references}

\appendix
\section{Appendix: Additional Experimental Details}
\label{app:additional-details}

\subsection{Routing and Readout Formulation}
\label{app:routing-readout}

This subsection provides the formal definitions of the routing mechanism
and readout weight computation referenced in Section~\ref{sec:method}.
The depth router makes a hard stop-or-continue decision in the forward
pass, while the readout retains a soft route mass through which the
recommendation loss can update the router. Let $d_i\in\{1,\ldots,R\}$
denote the hard selected depth of position $i$. Let $c_i^{(1)}=1$ denote
the route mass reaching the first boundary. For $r<d_i$, the continuation
mass reaching the next boundary is
\begin{equation*}
c_i^{(r+1)}
=c_i^{(r)}(1-q^{(r)}_{i,\mathrm{exit}}).
\end{equation*}
The readout weight at the selected depth is
\begin{equation*}
\omega_i=
\begin{cases}
c_i^{(d_i)}q^{(d_i)}_{i,\mathrm{exit}} & d_i<R,\\
c_i^{(R)} & d_i=R.
\end{cases}
\end{equation*}
For a fixed hard route, $\omega_i$ is differentiable with respect to the
evaluated router probabilities and provides a path for the recommendation
loss to update the router.

\paragraph{Balancing mass versus readout weight.}
The balancing mass $\alpha_{i,r}$ in the load-balancing term
(Section~\ref{sec:objective}) differs from the readout weight
$\omega_i$. While $\alpha_{i,r}$ distributes soft route mass over all
depths evaluated along the realized path to encourage balanced depth
usage, $\omega_i$ assigns soft mass only to the selected depth and
contributes to the final representation for the recommendation loss.

\paragraph{Routing warm-up.}
During routing warm-up, all valid positions are processed through the
full maximum recursion depth $R$. Routing decisions are bypassed and
router-side regularizers are disabled. The readout weight is set to
$\omega_i=1$ for every valid position.

\paragraph{Forward procedure.}
Algorithm~\ref{alg:looprec} summarizes adaptive recursion and
readout with routing enabled. The active set $\mathcal{A}$ contains
positions that have not exited, and $\mathcal{I}$ excludes padding.
For compactness, recursion superscripts are omitted from the current
states $\mathbf{H}$, exit probabilities $q_{i,\mathrm{exit}}$, and
continuation masses $c_i$.

\begin{algorithm}[t]
\caption{Adaptive Recursion and Readout in LoopRec}
\label{alg:looprec}
\SetAlFnt{\small}
\KwIn{Input representations $\mathbf{X}$, valid positions $\mathcal{I}$,
shared backbone $\mathcal{B}$, router $g$, recursion embeddings
$\{\boldsymbol{\rho}_r\}_{r=1}^{R-1}$, and maximum depth $R$}
\KwOut{Representations $\{\bar{\mathbf{z}}_i\}_{i\in\mathcal{I}}$
and selected depths $\{d_i\}_{i\in\mathcal{I}}$}

\tcp{Initial context encoding (recursion 1)}
$\mathbf{H}\leftarrow\mathcal{B}(\mathbf{X})$\;
$\mathcal{A}\leftarrow\mathcal{I}$; $c_i\leftarrow 1$ for all $i\in\mathcal{I}$\;

\BlankLine
\tcp{Depth-routed recursion}
\For{$r=1,\ldots,R-1$}{
    \tcp{Exit decisions at depth $r$}
    $q_{i,\mathrm{exit}}\leftarrow 1-\sigma(g(\mathbf{h}_i+\boldsymbol{\rho}_r))$ for all $i\in\mathcal{A}$\;
    $\mathcal{E}\leftarrow\{i\in\mathcal{A}:q_{i,\mathrm{exit}}\ge 0.5\}$\;
    $\omega_i\leftarrow c_i q_{i,\mathrm{exit}}$, $d_i\leftarrow r$ for all $i\in\mathcal{E}$\;
    $\mathcal{A}\leftarrow\mathcal{A}\setminus\mathcal{E}$\;
    \If{$\mathcal{A}=\varnothing$}{
        \textbf{break}\;
    }
    $c_i\leftarrow c_i(1-q_{i,\mathrm{exit}})$ for all $i\in\mathcal{A}$\;
    \tcp{Apply recursion $r+1$ to continuing positions}
    $\mathbf{H}_{\mathcal{A}}\leftarrow\mathcal{B}(\mathbf{H}_{\mathcal{A}})$\;
}

\BlankLine
\tcp{Forced exit at depth $R$ and readout}
$\omega_i\leftarrow c_i$, $d_i\leftarrow R$ for all $i\in\mathcal{A}$\;
$\bar{\mathbf{z}}_i\leftarrow\mathbf{x}_i+\omega_i\mathbf{h}_i$ for all $i\in\mathcal{I}$\;
\Return{$\{\bar{\mathbf{z}}_i,d_i\}_{i\in\mathcal{I}}$}\;
\end{algorithm}

The active subsequence $\mathbf{H}_{\mathcal{A}}$ preserves the
original chronological order under the causal mask. Exited positions
retain their states, so the final $\mathbf{h}_i$ is the state at the
selected depth $d_i$.

\subsection{Backbone-Specific Implementation}
\label{app:backbone-details}

We provide implementation details for applying LoopRec to
different backbones, including the recursive unit, the wrapping choice
around each original backbone, the main hyperparameter search spaces,
batch size, and backbone-specific hyperparameters.

We train all models using the Adam optimizer for up to 100 epochs, with early stopping patience of 10 based on validation NDCG@20. Test performance is evaluated using the checkpoint with the best validation score, and main results are averaged over five runs with different random seeds. The ``*'' in the main table denotes statistical significance ($p < 0.05$) for LoopRec versus the corresponding fixed-depth backbone under paired t-tests.
For the full LoopRec models, we fix the z-loss coefficient $\lambda_z$ to $10^{-3}$ across all datasets and backbones.

\paragraph{SASRec.}
The LoopRec recursive unit is a SASRec-style Transformer block. A
configuration \texttt{LxRy} applies $x$ blocks per recursion round
for up to $y$ rounds with adaptive depth routing. We search learning rate in
$\{3{\times}10^{-4},\ 6{\times}10^{-4},\ 10^{-3}\}$, encoder layers in $\{1,2,3,4,5,6\}$, L2 in $\{10^{-6},
10^{-4}\}$, dropout in $\{0.2,\ 0.5\}$, LoopRec configurations in
$\{\texttt{L1R4},\ \texttt{L2R2},\ \texttt{L1R6},\ \texttt{L2R3},
\ \texttt{L1R8},\ \texttt{L2R4}\}$, router warm-up
$W_{\mathrm{router}}\in\{5,8,10,12,16\}$, and balancing coefficient
in $\{0.001,\ 0.003,\ 0.005,\ 0.01\}$. The batch size is fixed to
$1024$.

\paragraph{BSARec.}
BSARec preserves the SASRec-family interface while adding a
frequency-domain inductive bias. We apply depth-routed recursion around
the native BSARec block, share the SASRec-family search ranges for
layers, learning rate, L2 regularization, dropout, LoopRec
configuration, $W_{\mathrm{router}}$, and balancing coefficient, and
additionally search frequency hyperparameters
$\alpha\in\{0.7,\ 0.9\}$ and $c\in\{3,\ 5\}$. The batch size is fixed
to $512$.

\paragraph{NextItNet.}
Recursion is applied at the native residual convolution block.
We fix the native encoder to six residual
blocks with a $(1,2,4)$ dilation cycle, calibrate learning rate within
this boundary, search batch size over $\{256,\ 512,\ 1024\}$, and
search LoopRec configurations in $\{\texttt{L3R2},\ \texttt{L3R3},
\ \texttt{L6R2}\}$.

\paragraph{LRURec.}
Recursion is applied at the native LRURec block. We search $1$--$3$
LRURec blocks, learning rate in $\{0.3,\ 0.6,\ 1,\ 3\}\times10^{-3}$,
and dropout in $\{0.4,\ 0.6\}$, with batch size fixed to $512$. We
otherwise share the L2 regularization, LoopRec configuration,
$W_{\mathrm{router}}$, and balancing ranges.

\begin{figure}[t]
  \centering
  \includegraphics[width=\linewidth]{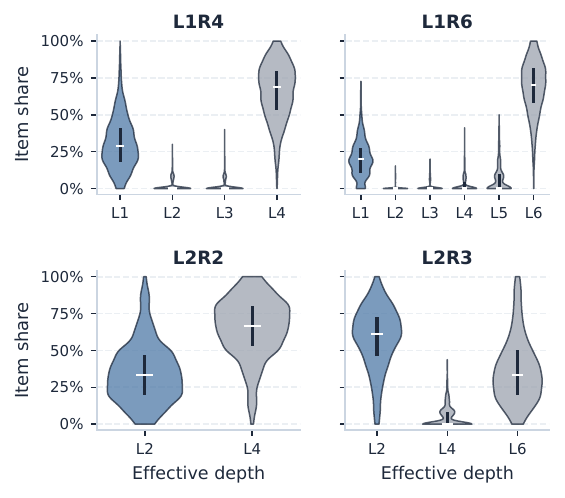}
  \caption{Selected-depth distributions on Office for four \texttt{LxRy} configurations. The y-axis is the per-sequence share of positions at each effective depth.}
  \label{fig:sasrec-office-depth-allocation}
\end{figure}

\subsection{Fixed-Depth Diagnostic Details}
\label{app:fixed-depth-diagnostics}

This subsection provides the experimental details for the fixed-depth
recommender diagnostics in Section~\ref{sec:rethinking}, including the
counterfactual truncation protocol for Top-30 exact-set saturation and
the candidate construction and tail-side computation constraint for the
restricted Tail-5 oracle.

\paragraph{Top-30 Exact-Set Saturation.}
For the truncation diagnostic in Section~\ref{sec:rethinking}, we use
a trained three-layer SASRec on Amazon Office Products. The main text
defines the saturation statistic $\ell_i^\star$. In the counterfactual
pass, only the selected history position $i$ stops being updated after
layer $\ell$. All other positions continue through the normal forward
pass. We use $\ell_i^\star$ purely as a stability statistic for the
trained model's full-depth recommendation set.

\paragraph{Restricted Tail-5 Oracle.}
For the oracle diagnostic, we keep the parameters of the trained
three-layer SASRec fixed on Amazon Office Products and vary only the
update depths of the five most recent history items. Earlier positions
follow the original full-depth forward pass. Each tail candidate assigns
one of the available depths in $\{1,2,3\}$ per tail position. After that
depth, the position's representation stops being updated for the
remaining layers. This yields a small candidate family over the five
most recent items. The original fixed-depth assignment is included as
one candidate, and we exclude assignments that use more tail-side layer
evaluations than this baseline. For each test sequence, the oracle
selects the candidate that gives the target item the best rank. Relative
tail computation is the number of layer evaluations on the five tail
items, normalized by the fixed-depth tail allocation.

\subsection{Depth Distribution Analysis}
\label{app:depth-distribution}

This subsection examines whether LoopRec's learned router
exploits multiple depths or collapses to a near-uniform strategy.
We analyze the distribution of selected depths across test sequences to
assess variation in computational allocation.

To quantify this, we compute the fraction of valid positions assigned to
each effective depth for every test sequence and summarize these
fractions across the test set. Figure~\ref{fig:sasrec-office-depth-allocation}
reports the resulting distribution for several LoopRec configurations,
denoted \texttt{LxRy}, where each recursion round applies $x$ shared
blocks and at most $y$ rounds are available to the router.

\noindent\textbf{LoopRec uses multiple effective depths.}
Figure~\ref{fig:sasrec-office-depth-allocation} shows that all four configurations use multiple effective depths, with substantial shares at shallow and deep depths. The non-uniform allocations indicate variation across positions and are consistent with the coexistence of shallow- and deep-saturating items in Section~\ref{sec:rethinking}.

\begin{table}[t]
\centering
\small
\setlength{\tabcolsep}{1.8pt}
\renewcommand{\arraystretch}{1.0}
\caption{Item-level routing of one Office sequence. Freq. is training-time interaction frequency. Depth is the selected depth.}
\label{tab:item-routing-case}
\begin{tabular*}{\columnwidth}{@{\extracolsep{\fill}}lp{0.40\columnwidth}p{0.24\columnwidth}rr@{}}
\toprule
\textbf{Pos.} & \textbf{Item} & \textbf{Category}
& \textbf{Freq.} & \textbf{Depth} \\
\midrule
1 & ValueMailers poly mailers & Envelope mailer & 56 & 1 \\
2 & Uboxes poly mailers & Envelope mailer & 261 & 1 \\
3 & ValueMailers poly mailers & Envelope mailer & 517 & 1 \\
4 & EasyPAG desk organizer & Desk organizer & 5 & 1 \\
5 & Quartet whiteboard & Dry erase board & 259 & 1 \\
6 & Aviditi box mailers & Box mailer & 1 & 2 \\
7 & PACKZON shipping labels & Printer label & 34 & 4 \\
8 & Hygloss library pockets & Teaching material & 4 & 3 \\
9 & MFLABEL address labels & Address label & 30 & 1 \\
10 & Avery address labels & Address label & 1,413 & 1 \\
11 & Pro Supply poly mailers & Envelope mailer & 1 & 3 \\
12 & Springhill cardstock & Colored paper & 17 & 1 \\
13 & TECKNET mouse pad & Mouse pad & 86 & 4 \\
target & Flash Furniture office chair & Office chair & -- & -- \\
\bottomrule
\end{tabular*}
\end{table}

\subsection{Item-Level Routing Examples}
\label{app:item-routing-case}

To complement the aggregate statistics in Section~\ref{sec:analysis-item-signals}, this subsection presents a concrete sequence example showing how LoopRec's router assigns depths to individual items in context. We also provide a category-level aggregation to illustrate the frequency-depth relationship at a coarser granularity.

Table~\ref{tab:item-routing-case} gives a full sequence from the Office
\texttt{L1R4} analysis in Section~\ref{sec:analysis-item-signals}. The
example reflects the two aggregate tendencies discussed in the main
analysis: later positions are more likely to receive deeper updates, and
high-frequency items tend to stay shallow. For instance, frequent mailer
and label items such as \textit{Uboxes poly mailers}, \textit{ValueMailers
poly mailers}, and \textit{Avery address labels} exit at depth 1, while
rarer items such as \textit{Aviditi box mailers}, \textit{Hygloss library
pockets}, and \textit{Pro Supply poly mailers} receive depths 2 or 3. The
exact selected depth is not determined by either signal alone:
\textit{PACKZON shipping labels} and \textit{TECKNET mouse pad} reach
depth 4, whereas the later \textit{Springhill cardstock} item remains
shallow, indicating that recency, popularity, category semantics, and
surrounding context jointly shape item-level routing.

\begin{figure}[t]
  \centering
  \includegraphics[width=\linewidth]{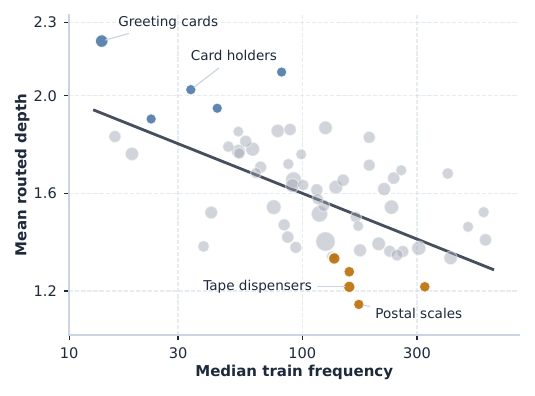}
  \caption{Category-level routed depth versus item popularity on Office. Each point is a category leaf (size $\propto$ token count). Blue and orange mark the five highest- and lowest-depth categories, and the solid line shows the overall trend.}
  \label{fig:category-depth-popularity}
\end{figure}

\paragraph{Category-level aggregation.}
We join routed history tokens with Amazon Reviews 2023 metadata and group
them by category leaf. Figure~\ref{fig:category-depth-popularity} plots
each category by its median training frequency and mean routed depth.
Categories with larger median frequency tend to have lower mean depth,
consistent with the item-frequency trend in
Table~\ref{tab:item-popularity-depth}. Individual routing decisions still
vary by position and context within each category.

\subsection{Comparison with Heuristic Routing}
\label{app:heuristic-routing}

We compare LoopRec with two heuristic routing strategies to
examine whether simple item signals can provide effective depth
allocation. We conduct this comparison on Office using the SASRec
backbone.

\paragraph{Heuristic routing.}
Both variants retain LoopRec's shared recursive backbone and
replace the learned router with a position- or frequency-based rule.
Each strategy assigns a score $s_i\in[0,1]$ to each valid position
$i$ and maps it to recursion rounds via
\begin{equation*}
r_i=1+\left\lfloor (R-1)s_i+\tfrac12\right\rfloor,
\end{equation*}
where $R$ is the maximum number of rounds. Larger scores assign
more recursion rounds.

\textbf{Position} assigns greater depth to more recent
positions. We index the $n$ valid items in the current input window from
earliest to most recent. For $n>1$, the score is $s_i=(i-1)/(n-1)$.
For $n=1$, we set $s_i=1$, treating the sole item as the most recent
position and assigning it maximum depth.

\textbf{Frequency} assigns greater depth to less frequent
items. We compute each item's interaction count $f(v)$ from the
training split, apply $g(v)=\log(1+f(v))$, and define
\begin{equation*}
s_i=1-\frac{g(v_i)-g_{\min}}{g_{\max}-g_{\min}},
\end{equation*}
where the minimum and maximum are computed over items observed
in training. Training interactions are counted once, without duplicating
occurrences across overlapping training examples.

Padding positions are excluded from routing. Both variants
retain the full-depth training warm-up and use their heuristic rules
throughout evaluation. The heuristic routes use unit
readout weights and omit router-specific regularization.

\paragraph{Experimental setup.}
We follow the same implementation details as the SASRec-based experiments in Table~\ref{tab:main-results}. All models are trained for up to 100 epochs with early stopping, and results are averaged over five runs with different random seeds.

\begin{table}[t]
\centering
\small
\setlength{\tabcolsep}{1.8pt}
\renewcommand{\arraystretch}{1.0}
\caption{Comparison with heuristic routing on Office using
SASRec. Values are test means over five seeds. \textbf{Bold} values denote
the highest mean for each metric.}
\label{tab:heuristic-routing}
\begin{tabular*}{\columnwidth}{@{\extracolsep{\fill}}lrrrr@{}}
\toprule
\textbf{Method} & \textbf{MRR@10} & \textbf{MRR@20}
& \textbf{NDCG@10} & \textbf{NDCG@20} \\
\midrule
\textsc{Vanilla} & 0.0508 & 0.0513 & 0.0562 & 0.0582 \\
Position & 0.0508 & 0.0515 & 0.0558 & 0.0585 \\
Frequency & 0.0510 & 0.0516 & 0.0561 & 0.0584 \\
LoopRec & \textbf{0.0523} & \textbf{0.0530}
& \textbf{0.0571} & \textbf{0.0595} \\
\bottomrule
\end{tabular*}
\end{table}

\begingroup
\setlength{\emergencystretch}{1em}
\paragraph{Results.}
Table~\ref{tab:heuristic-routing} compares ranking quality without matching average computation. Both heuristic strategies yield modest gains in NDCG@20 over vanilla SASRec. These gains are consistent with the trends observed in Sections~\ref{sec:analysis-position-depth} and~\ref{sec:analysis-item-signals}, where more recent positions and less frequent items tend to receive deeper computation. Although the
gains do not extend to every metric, they provide complementary evidence
that position and frequency can serve as useful signals for depth
allocation. LoopRec achieves the highest mean on all four metrics,
with NDCG@20 exceeding Position and Frequency by 1.64\% and 1.88\%,
respectively.
\par
\endgroup

\subsection{Runtime and Latency Measurements}
\label{app:runtime-latency}

LoopRec's adaptive routing reduces the computational budget by skipping
layers for shallow-exiting positions. This subsection provides encoder
latency measurements for several LoopRec configurations alongside their
computational savings.

\begin{table}[h]
\centering
\small
\setlength{\tabcolsep}{1.4pt}
\caption{Compute budget and encoder latency on Office (single A6000 GPU). \textbf{Compute} and \textbf{Latency} are normalized by the matched fixed-depth reference. Ref.\,ms and LoopRec\,ms are median encoder times.}
\label{tab:runtime-budget-latency}
\begin{tabular*}{\columnwidth}{@{\hspace{\tabcolsep}\extracolsep{\fill}}llrrrrr@{\hspace{\tabcolsep}}}
\toprule
\textbf{Setting} & \textbf{Ref.} & \textbf{Compute}
& \textbf{Ref.\,ms} & \textbf{LoopRec\,ms} & \textbf{$\Delta$ms}
& \textbf{Latency} \\
\midrule
\texttt{L2R4} & SASRec-8L & 0.407$\times$ & 91.230 & 79.273 & $-11.957$ & 0.869$\times$ \\
\texttt{L2R3} & SASRec-6L & 0.680$\times$ & 69.684 & 63.196 & $-6.488$ & 0.907$\times$ \\
\texttt{L2R2} & SASRec-4L & 0.755$\times$ & 47.624 & 47.379 & $-0.245$ & 0.995$\times$ \\
\bottomrule
\end{tabular*}
\end{table}

We report \emph{encoder-only} latency rather than full-scoring latency,
because adaptive depth directly modifies the sequential encoder.
Table~\ref{tab:runtime-budget-latency} compares three configurations
against fixed-depth SASRec references at the same maximum effective depth.
All numbers are measured on a single NVIDIA A6000 GPU.

\noindent\textbf{Adaptive routing reduces compute, with setting-dependent latency gains.}
LoopRec reduces selected computation in all three settings. Encoder latency decreases for \texttt{L2R4} and \texttt{L2R3}, but remains nearly unchanged for \texttt{L2R2}. Compute savings thus only partly translate into wall-clock gains, depending on how dynamic execution maps onto the GPU runtime.

\end{document}